\documentclass[sigconf]{acmart}

\usepackage{balance}

\copyrightyear{2026}
\acmYear{2026}
\setcopyright{cc}
\setcctype{by}
\acmConference[UbiComp Companion '26]{Companion of the 2026 ACM International Joint Conference on Pervasive and Ubiquitous Computing}{October 11--15, 2026}{Shanghai, China}
\acmBooktitle{Companion of the 2026 ACM International Joint Conference on Pervasive and Ubiquitous Computing (UbiComp Companion '26), October 11--15, 2026, Shanghai, China}
\acmDOI{10.1145/3798063.3838813}
\acmISBN{979-8-4007-2533-3/2026/10}

\begin{document}

\title{Health Inquiry with AI: How Empathetic Expression and Conversational Contexts Shape Users' Communicative Acts}
\renewcommand{\shorttitle}{Users' Communicative Acts in Health Inquiry With AI}
\author{Xi Zheng}
\affiliation{%
  \institution{Department of Computer Science}
  \institution{City University of Hong Kong}
  \city{Hong Kong SAR}
  \country{China}
}
\email{zheng.xi@my.cityu.edu.hk}

\author{Xuyu Yang}
\affiliation{%
  \institution{School of Creative Media}
  \institution{City University of Hong Kong}
  \city{Hong Kong SAR}
  \country{China}
}
\email{xuyuyang2-c@my.cityu.edu.hk}

\author{Can Liu}
\affiliation{%
  \institution{School of Creative Media}
  \institution{City University of Hong Kong}
  \city{Hong Kong SAR}
  \country{China}
}
\email{canliu@cityu.edu.hk}

\author{Yuhan Luo}
\affiliation{%
  \institution{Department of Computer Science}
  \institution{City University of Hong Kong}
  \city{Hong Kong SAR}
  \country{China}
}
\email{yuhanluo@cityu.edu.hk}
\authornote{Corresponding Author.}
\renewcommand{\shortauthors}{Xi Zheng, Xuyu Yang, Can Liu, Yuhan Luo}




\begin{teaserfigure}
\centering
   \includegraphics[width=0.90\textwidth, height=0.37\textheight]{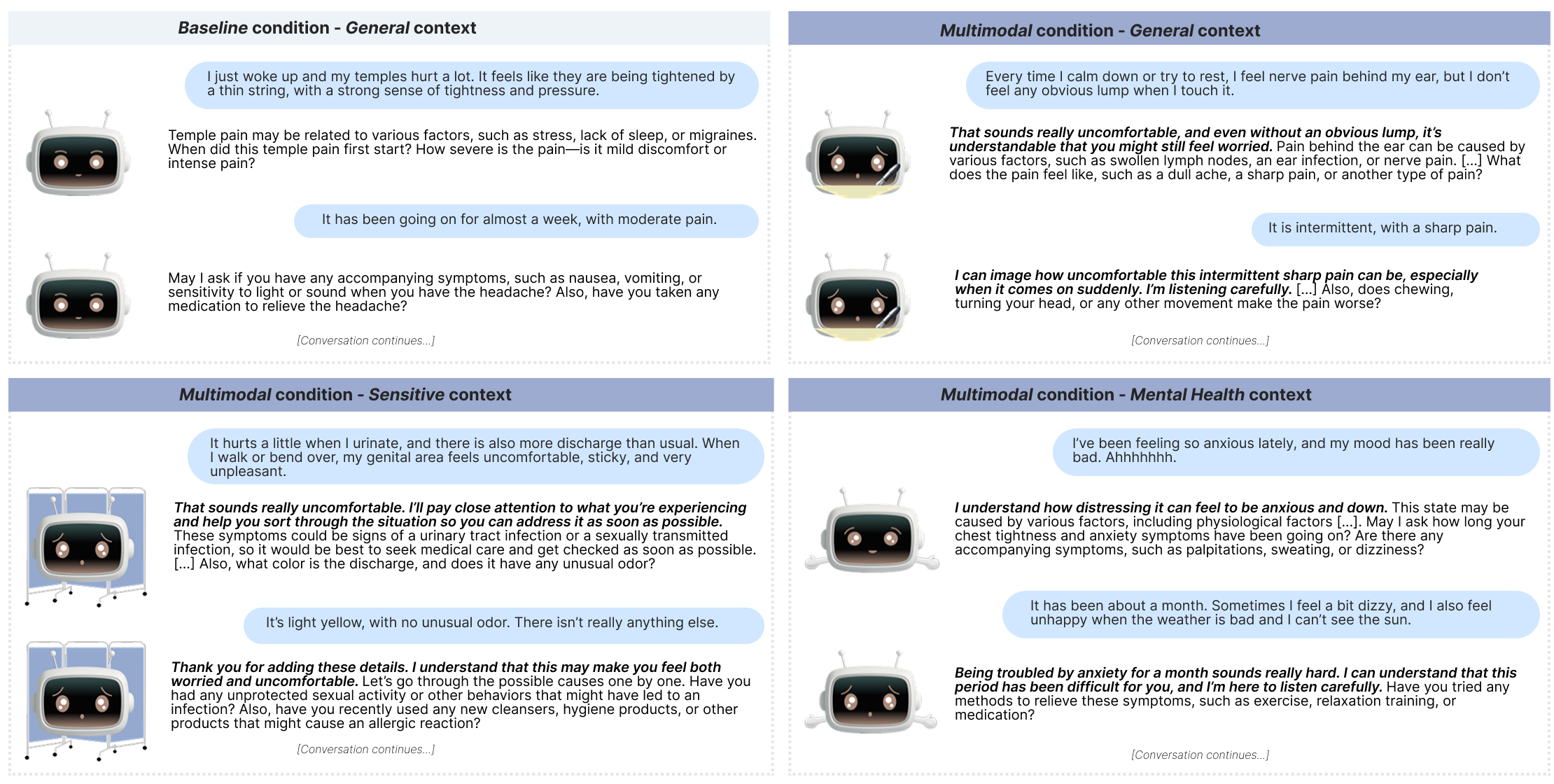}
    \caption{Example conversations across modality and conversational contexts. 
    Empathetic verbal responses are bold-italicized. 
    }
    \label{fig:examples}
  \Description{No description.}
  \label{fig:teaser}
\end{teaserfigure}

\begin{abstract}
As online health information-seeking shifts to conversational AI, high-quality information retrieval increasingly relies on users' ``communicative acts''(proactively sharing and seeking information)---similar to how effective diagnosis and personalized guidance are elicited in patient-clinician communication. Drawing on health communication research, this study examines how a chatbot's modality of empathetic expression (Verbal, Visual, Multimodal) and the conversational context (General, Sensitive, Mental Health) influence these acts through a $2 \times 2 \times 3$ within-subjects experiment ($N = 48$).
The results revealed that while verbal and multimodal empathy significantly increased reply length, communicative acts were largely shaped by conversational context, with Sensitive context triggering more question-asking and Mental Health context leading to heightened concerns, assertive responses, and unprompted information disclosure. 
Combined with qualitative findings, we discuss design implications for building context-sensitive AI health inquiry systems that can encourage active user participation.
\end{abstract}

\maketitle

\section{Introduction and Backgrounds}
Driven by the rapid evolution of large language models (LLMs), online information-seeking, particularly health inquires, is shifting away from traditional search engines to conversational AI~\cite{yun2025online, zhou2026understanding, abbasian2024personalized}.
Yet, the quality of information retrieved from AI depends not only on how well the model is designed and trained~\cite{singhal2023large, he2024quality}, but also individual's ability to communicate with (or prompt) the AI, such as asking follow-up questions or adding contextual details about their health conditions~\cite{kusa2023drprompt, rebitschek2025evaluating, ng2025prompt, liu2024using}.
This is similar to effective communication with human clinicians, where the patient's active inquiry and disclosure are essential for the clinician to fully understand their needs and concerns, and deliver more appropriate, personalized guidance~\cite{Street2001AnalyzingParticipation, cegala2009impact, d2017promoting}.

Health communication researchers have characterized this active participation as ``\textit{\textbf{communicative acts}},'' which include asking questions, expressing concerns, and making assertive responses ~\cite{Street2001AnalyzingParticipation}. These acts allow patients to articulate their needs, worries, preferences, and interpretations during clinical encounters~\cite{Street2001AnalyzingParticipation}.
For example, asking questions signals patient's knowledge gap; expressing concerns reveals their emotional distress and worries; and assertive responses clarifies their beliefs and treatment  expectations~\cite{Street2001AnalyzingParticipation, d2017promoting}. 
Prior work has shown that patients who engage in these communicative acts are more likely to elicit informative and accommodating responses, as these acts provide clinicians with cues needed to tailor explanations and decision-making support~\cite{Street2005PatientParticipation, d2017promoting, cegala2009impact}.

According to existing research in health communication, individuals' active participation in their inquiries is influenced not only by their own health literacy and communication skills~\cite{katz2007patient, Street2005PatientParticipation, d2017promoting}, but also the clinicians' communication styles and the nature of the inquiry topics~\cite{Street2005PatientParticipation, hashim2017patient, sankar2005tell, parker2020patients}. For example, when clinicians express empathy through verbal acknowledgment (e.g., ``\textit{I understand how pain it is}''~\cite{zhang2024customer, Seitz2024Empathy}) or supportive facial expressions (e.g., attentive nodding, a comforting smile~\cite{bavelas2000visible}), patients feel validated and emotionally safe, and thus are willing to ask more questions and openly express worries~\cite{hashim2017patient, Street2005PatientParticipation, d2017promoting}.
On the other hand, prior work has shown that patients' communicative acts and needs vary across clinical contexts. For example, sensitive health topics may make patients hesitant to disclose information because of stigma, embarrassment, and privacy concerns~\cite{sankar2005tell}. In contrast, patients with mental health concerns may want to seek opportunities for disclosing their emotional distress and gaining validation and supports~\cite{parker2020patients}.

Extending these findings from human-human interaction to human-AI interaction, we see the opportunities to encourage communicative acts through optimizing the design of AI systems for health inquiry~\cite{yue2023beyond, Stal2021faicial, zheng2026disentangling, liu2018should}. In this regard, we explore how the AI's \textbf{\textit{modality of empathetic expression}} and the \textbf{\textit{conversational context}} influence individuals' communicative acts.
Specifically, we examine three common health inquiry contexts: \textit{General}, \textit{Sensitive}, and \textit{Mental Health}, and design four conditions where the chatbot displayed empathy through different modality configurations: \textit{Baseline} (no explicit empathy cues), \textit{Verbal empathy only}, \textit{Visual empathy only}, and \textit{Multimodal empathy} (both verbal and visual cues) conditions.
By comparing participants' communicative acts across these conditions and contexts, we found that empathic verbal expressions increased participants' reply length, whereas the specific communicative acts users performed were more strongly shaped by the conversational context. 

This work contributes to HCI and the IMWUT community by providing an empirical understanding of user engagement with health inquiry AI through the lens of communicative acts. Our findings offer design implications for building context-sensitive AI systems for health inquiry to better elicit users' active information sharing and seeking behaviors that can result in higher quality of retrieved information.
\vspace{-2mm}

\section{Method}

\subsection{Data Source}
We analyzed conversation logs collected from a previously conducted 2 (\textit{without} vs. \textit{with verbal empathy}) $\times$ 2 (\textit{without} vs. \textit{with visual empathy}) $\times$ 3 (\textit{general, sensitive, mental health context}) within-subjects experiment ($N = 48$), followed by optional debriefing interviews ($n = 16$). The original experiment examined the effects of empathetic expression of health inquiry chatbot on people's empathy and authenticity perception, as well as their trust, satisfaction, and dependency toward the chatbot~\cite{zheng2026disentangling}. 
This paper reports a complementary analysis of the same dataset, shifting the analytical focus to users' interactions with the chatbot. Particularly, we compare their communicative acts across study conditions.

To maintain topic consistency and protect participant privacy, the health contexts were presented as hypothetical scenarios with visual illustrations and text descriptions rather than prompting users to discuss their actual medical conditions. 
This approach has been widely used in prior work to examine user perceptions in sensitive settings~\cite{liu2018should, yue2023beyond, genc2024empathyreview}.

Through a custom-built interface, participants first read a hypothetical scenario, adopted the scenario character's perspective, chatted with the chatbot about the described health issue, completed post-task ratings, and then proceeded to the next scenario.
The chatbot's responses were generated to manipulate whether empathy was expressed through verbal cues, visual cues, both channels, or neither.
The presentation of these conditions were counterbalanced and the design of empathetic expression passed manipulation check. Full details regarding the experimental setup, procedure, and the design of the empathetic cues are available in~\cite{zheng2026disentangling}. The experiment was conducted in mainland China and was approved by the ethics review committee in the authors' institution.
Figure~\ref{fig:examples} illustrates the conversation examples collected from the experiment. 

\vspace{-2mm}

\subsection{Data Analysis}

To understand participants' reactions to different empathy modalities, we analyzed only the dialogue occurring after the chatbot's first reply (after participants were exposed to the condition). 
We first coded each of the messages based on Street's definition of communicative acts~\cite{Street2001AnalyzingParticipation}, and then compared their likelihood to appear across conditions. 
Next, we triangulate these quantitative data with participants' qualitative pre- and post-task responses and interviews to contextualize the findings. 

\subsubsection{Communicative Acts Coding}

According to Street et al., communicative acts---the process of active patient participation during encounters with clinicians---include \textit{asking questions}, \textit{expressing concerns}, and \textit{making assertive responses}, which are essential for effectively addressing one's informational needs~\cite{Street2001AnalyzingParticipation}. Additionally, we extended this framework by incorporating \textit{providing unprompted contextual information}, to capture the proactive sharing of personal background or symptom history that complements the picture of their health condition and concerns~\cite{d2017promoting, Pereira2023Whyjohnny}. 

Two researchers independently coded the same one-third of the dataset based on the four dimensions of communicative acts, and then compared their coding results to discuss and resolve any discrepancies. Next, they independently coded an additional one-third of the dataset to achieve an inter-coder reliability of Cohen's $\kappa = .94$, then the first author finished coding the remaining data. 
In the following, we describe each of these dimensions in detail with example quotes from participants' message. Note that one message may contain multiple communicative acts.

\begin{itemize}
    \item \textbf{Asking questions}:  seeking information, clarification, or guidance in an interrogative form (e.g., ``\textit{sometimes I feel a little dizzy. Does that count as another symptom?}'' and ``\textit{I haven't showered for several days. Could that be the main reason, or is it something related to my body itself?}'').

    \item \textbf{Expressing concerns}: expressing worry, anxiety, fear, frustration, anger, or other negative emotions (e.g., ``\textit{What could be causing this? I'm actually quite scared right now.}'' and ``\textit{Fatigue has been constant. You can probably tell that I really have too much on my plate.}'').

    \item \textbf{Making assertive responses}: stating one's beliefs, preferences, interpretations, or expectations of the condition, which may involve disagreement and suggestions (e.g., ``\textit{I did not use medication to help me sleep, because I think that if I can fall asleep on my own, I should avoid relying on medication.}'' and ``\textit{I don't really think it's likely to be an infection? It was just contact with unclean underwear.}'').

    \item \textbf{Providing unprompted contextual information}: mentioning contexts about one's situations that are not included in the scenario descriptions of the study materials or asked by the chatbot, which sometimes involved participants' own experience or expectations 
    (e.g., ``\textit{Recently the weather has been gloomy, and I have been feeling very down. I do not want to do anything, my chest feels tight, and I keep feeling anxious.}'' and ``\textit{No, I don't want to exercise because I feel like I don't have much energy in my body.}'').
\end{itemize}



\subsubsection{Quantitative Analysis}
We first compared participants' reply length (total number of Chinese characters) per message using a linear mixed-effects model~\cite{RLanguage}.
The model included the experimental condition and their interaction as fixed effects, with each participant as a random effect. 
Next, we analyzed the above coded communicative acts separately using mixed-effects logistic regression, with the same fixed- and random-effects structure. Finally, we conducted robustness checks to assess the stability of the results.
To validate the trends observed about the communicative acts across conversational contexts, we conducted an independent analysis of participants' initial messages sent to the chatbot (upon seeing the conversational context and before exposing to the empathy modality) and verified the results were consistent.
\vspace{-2mm}
\section{Results}
\begin{figure*}[t]
    \centering
    \includegraphics[width=0.93\textwidth, height=0.35\textheight]{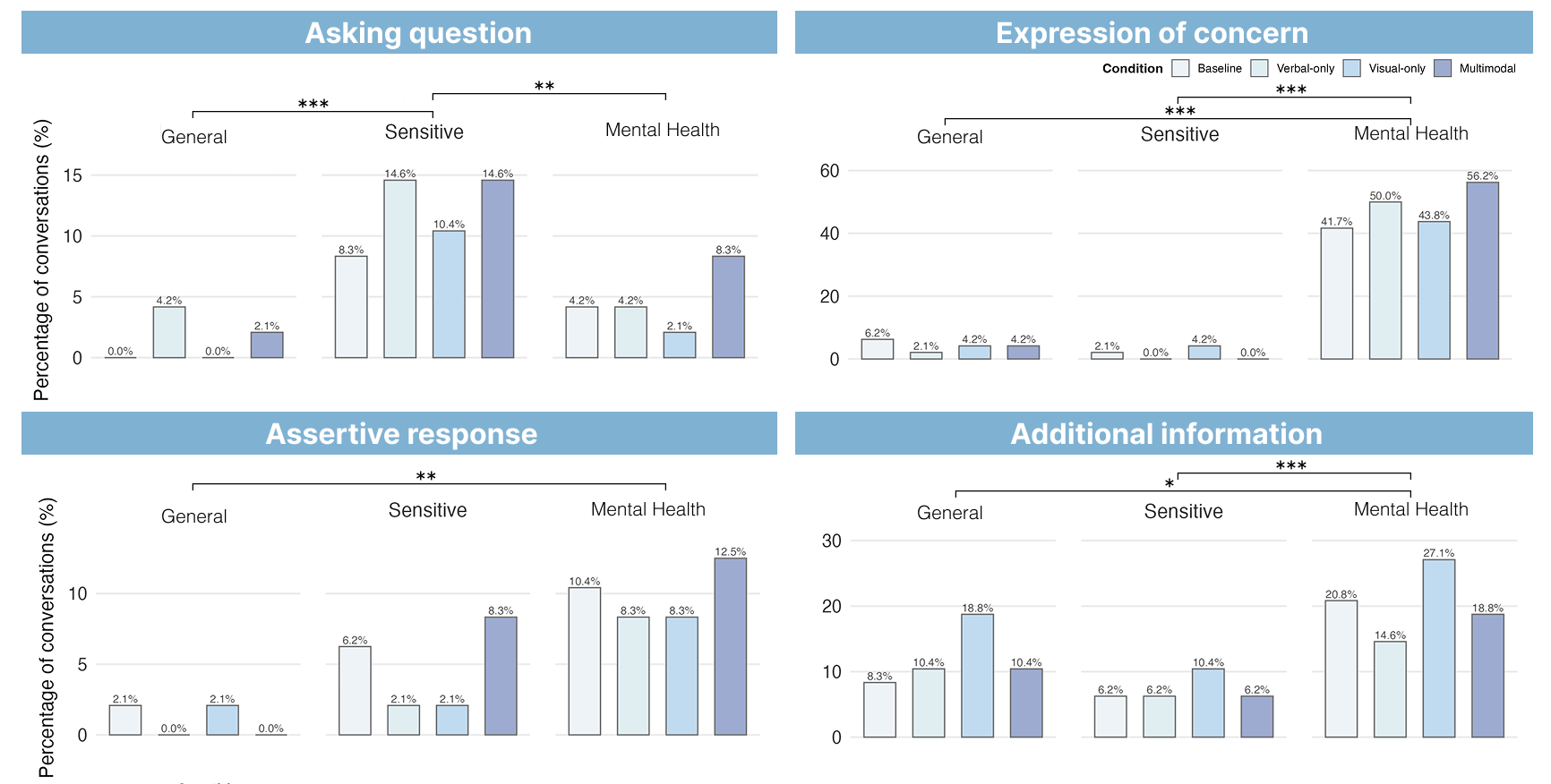}
    \caption{Main result visualization of users' communicative acts across empathy conditions and health contexts. Statistically significant differences across conversational contexts are marked. Note. * $p < .05$, ** $p < .01$, *** $p < .001$.}
    \label{fig:main_results}
\end{figure*}

\subsection{Reply Length}
Compared to the \textit{Baseline} condition, participants typed significantly longer replies (5--6 more characters on average) in the \textit{Verbal-only} ($\beta = 5.54$, $p = .024$) and \textit{Multimodal} conditions ($\beta = 5.60$, $p = .024$). No significant effect was found between the \textit{Visual-only} and the \textit{Baseline} condition($\beta = 2.01$, $p = .598$). Regarding conversational contexts, replies were significantly longer in the \textit{Mental Health} context than in the \textit{General} ($\beta = 7.27$, $p < .001$) and \textit{Sensitive} contexts ($\beta = 6.06$, $p < .001$). The \textit{General} and \textit{Sensitive} contexts did not significantly differ from each other ($\beta = 1.21$, $p = .476$). No significant interaction effect was identified between empathy cues and health contexts on reply length ($\chi^2(6) = 0.50$, $p = .998$).
\vspace{-2mm}
\subsection{Communicative Acts}
Across the four communicative acts---asking questions, expressing concerns, making assertive responses, and providing unprompted contextual information---most measures significantly differed across health contexts, while only a few minor differences were identified across empathy modality conditions. 
Specifically, none of the contrasts in empathy modality reached statistical significance, despite that the \textit{Verbal-only} ($\Delta = 0.96$, $p = .5407$) and \textit{Multimodal} conditions ($\Delta = 1.10$, $p = .4918$) showed higher rates of question-asking than the \textit{Baseline} condition and the \textit{Visual-only} condition showed a higher likelihood of providing unprompted contextual information than \textit{Baseline} ($\Delta = 0.70$, $p = .3325$).
Below, we focused on describing the varied interaction patterns across conversational contexts.
\vspace{-2mm}
\subsubsection{More Question Asking in \textit{Sensitive} Context} Participants were more likely to ask questions in \textit{Sensitive} context than in both the \textit{General} context ($\Delta = 2.9$, $p = .0002$) and the \textit{Mental Health} context ($\Delta = 1.5$, $p = .0068$).
From their conversation logs and interpretations noted in the scenarios, we found that in the \textit{Sensitive} context, participants were navigating a tense intersection of social stigma and acute anxiety, noting that ``\textit{This symptom feels really embarrassing}'' and sometimes raising multiple questions at once: ``\textit{Could this have been caused by not changing my underwear, or cleaning my private area over the past few days?}''  
As a result, they exhibited an urgent, protective need to immediately identify the exact cause and resolve the issue (e.g., ``\textit{What is happening? What exactly are these small bumps? How can they go away?}'')
\vspace{-2mm}
\subsubsection{Heightened Concerns, More Assertive Responses and Disclosure in \textit{Mental Health} Inquiries}
In \textit{Mental Health} context, participants expressed higher intensity of concerns than in \textit{General} ($\Delta = 3.07$, $p < .0001$) and \textit{Sensitive} contexts ($\Delta = 4.08$, $p < .0001$); they were also more likely to input assertive responses ($\Delta = 2.80$ vs. \textit{General}, $p = .0018$) and provide unprompted contextual information (\textit{General}: $\Delta = 0.79$, $p = .0284$; \textit{Sensitive}: $\Delta = 1.45$, $p = .0003$). 
The qualitative data reveals that because mental health struggles lack visible biomarkers, participants relied heavily on narrative disclosure, situating their distress in relation to its duration, triggers, and daily impacts to make their internal states understandable to the AI. This included keyword labeling of their conditions such as \textit{``anxious,''} \textit{``frustrated,''} \textit{``helpless,''} and \textit{``exhausted,''} and sometimes ``self-diagnosis'' that attributed their conditions to ``\textit{stress}'', ``\textit{depression}'', ``\textit{ADHD}'', or ``\textit{difficulties with emotion regulation}.'' Moreover, to convey the severity of their concerns, they shared unprompted details on how their sleep, work performance, relationships, and daily functioning are impacted.

\vspace{-5mm}
\section{Discussion}

Our analysis showed that while the modality of empathetic expression increased participants reply length, such increase did not necessarily contribute to communicative acts. This finding suggested that while empathetic cues (particularly verbal and multimodal ones) are effective at keeping users engaged and encouraging them to type more, the additional text may consist of conversational politeness or social reciprocity~\cite{BOWMAN2024politeness, GUO2025empathic} rather than proactive information-seeking or assertive behaviors.
On the other hand, we found that participants' communicative acts were mainly driven by the conversational context---the complexity and urgency of their health conditions. This indicates that active participation, such as intensive information-seeking, sharing one's diagnostic interpretations, and volunteering unprompted context, is fundamentally a problem-driven coping mechanism rather than a reaction to the AI's social behaviors~\cite{Street2005PatientParticipation, mcmullan2019relationships, rains2015information}.

Taken together, we see the opportunities to tailor the AI's communication strategy and inquiry interface based on the conversational contexts, so as to maximize the benefits of the communicative acts that users exhibit. For example, when confronted with the acute anxiety and stigma of a \textit{Sensitive} condition, the system can deploy structural UI scaffolding, such as interactive FAQ accordion blocks~\cite{sansoni2015question, tracy2022question} or etiology-tracing visualizations to rapidly provide cognitive closure~\cite{woodcock2021impact}.
When navigating \textit{Mental Health} burdens, the AI may prioritize understanding and compassion to alleviate their concerns~\cite{parker2020patients}. This sheds light on combining empathetic expression of AI; although in our findings, the effects of empathy modality on communicative acts were not significant, it was still essential for keeping participants engaged.
At the same time, how to dynamically process the information from users' self-disclosure and handle their self-diagnosis without over-stepping into clinical diagnosis or dismissively shutting down their intuition remains an open challenge for future research~\cite{huo2025large}.

\section{Limitations and Future Work}

To maintain consistent experimental conditions, we held the AI responses constant; as a result, our study does not establish whether the characterized ``communicative acts'' improve the model's response quality~\cite{du-etal-2025-context, shi2023distract}. Future work should involve clinical professionals in assessing AI responses to these communicative acts with real-world conversation logs.
In addition, the results of this study may not generalize across cultures~\cite{Sheeran2023culture,Mojaverian2013culture}.

\begin{acks}
We thank our participants for their time and contributions. We also appreciate the thoughtful suggestions provided by the anonymous reviewers.
This work is supported by City University of Hong Kong (\#7005997).
\end{acks}

\bibliographystyle{ACM-Reference-Format}
\balance{}
\bibliography{sample-base}

\appendix

\end{document}